\documentclass[nofootinbib,aps,superscriptaddress]{revtex4}
\usepackage{amssymb,amsmath,graphicx,color}
\begin{document}

\preprint{IPMU13-0240}

\title{A no-go theorem for generalized vector Galileons on flat spacetime}

\author{C\'edric Deffayet}
\affiliation{UPMC-CNRS, UMR7095, Institut d'Astrophysique de Paris, GReCO, 98bis boulevard Arago, F-75014 Paris, France}
\affiliation{IHES, Le Bois-Marie, 35 route de Chartres, F-91440 Bures-sur-Yvette, France}

\author{A. Emir G\"umr\"uk\c{c}\"uo\u{g}lu}
\affiliation{School of Mathematical Sciences, University of Nottingham, University Park, Nottingham, NG7 2RD, UK}
\affiliation{Kavli Institute for the Physics and Mathematics of the Universe, Todai Institutes for Advanced Study, University of Tokyo (WPI), 5-1-5 Kashiwanoha, Kashiwa, Chiba 277-8583, Japan}

\author{Shinji Mukohyama}
\affiliation{Kavli Institute for the Physics and Mathematics of the Universe, Todai Institutes for Advanced Study, University of Tokyo (WPI), 5-1-5 Kashiwanoha, Kashiwa, Chiba 277-8583, Japan}

\author{Yi Wang}
\affiliation{Centre for Theoretical Cosmology, DAMTP, University of Cambridge, Cambridge CB3 0WA, UK}
\affiliation{Kavli Institute for the Physics and Mathematics of the Universe, Todai Institutes for Advanced Study, University of Tokyo (WPI), 5-1-5 Kashiwanoha, Kashiwa, Chiba 277-8583, Japan}

\begin{abstract}
 We explore the possibility for generalized electromagnetism on flat
 spacetime. For a single copy of $U(1)$ gauge theory, we show that the
 Galileon-type generalization of electromagnetism is forbidden. Given
 that the equations of motion for the vector field are gauge invariant
 and Lorentz invariant, follow from an action and contain no more than
 second derivative on $A_\mu$, the equations of motion are at most
 linear with respect to second derivative of $A_\mu$. 
\end{abstract}

\maketitle

\newcommand{\ba}{\begin{eqnarray}}
\newcommand{\ea}{\end{eqnarray}}
\newcommand{\ndd}{n}
\newcommand{\npi}{N}

\section{Introduction}
Scalar fields models with derivative self-interactions and second order
field equations have attracted considerable attention recently. Inspired
from the decoupling limit of the Dvali-Gabadadze-Porrati (DGP) model
\cite{Dvali:2000hr}, scalar field models known as Galileons have been
introduced on flat spacetime \cite{Nicolis:2008in}. These Galileons can
be defined as the most general Lorentz invariant scalar models which
have field equations containing only second order derivatives on
Minkowski spacetime. In four dimensions, there are only four nontrivial
such theories\footnote{Five, if one includes the tadpole.}. Galileons
have been extended to curved spacetime
\cite{Deffayet:2009wt,Deffayet:2009mn}, and to a larger family covering
all the (single) scalar field models which have field equations of order
less than or equal to two on flat spacetime (as opposed to the original
Galileons which have field equations of exactly second order)
\cite{Deffayet:2011gz}. In fact, those models have been discovered much
earlier than the introduction of the Galileons, first by Horndeski, who
obtained all the scalar-tensor models in four dimensions which have
field equations (for both the metric and the scalar field) of order less
than or equal to two \cite{Horndeski:1974wa}, and later by Fairlie {\it
et al.} in a different context
\cite{Fairlie:1992nb,Fairlie:1992he,Fairlie:1991qe} using ``Euler
hierarchies''.  Such models (for a review see
e.g. \cite{Deffayet:2013lga}) have been used in a variety of situations,
including cosmology. A natural question to ask is whether models with
similar properties can be built for fields of higher
spins. Ref. \cite{Deffayet:2010zh} has shown how to obtain the
equivalent for $p$-forms of the Galileons and their covariantization,
and also presented a natural generalization to the case of multi
$p$-forms. Such ``multi $p$-forms generalized Galileons'' include in
particular multi-scalar Galileons, or ``multigalileons'' which were
later studied
\cite{Padilla:2010de,Padilla:2010tj,Hinterbichler:2010xn,Trodden:2011xh,Sivanesan:2013tba,Padilla:2012dx,Kobayashi:2013ina}. 
However, the construction in \cite{Deffayet:2010zh}, when restricted to
theories containing a single $p$-form, leads only to trivial
(i.e. theories with vanishing field equations on flat spacetime)
theories whenever $p$ is odd and regardless of the spacetime
dimension. And indeed, e.g., there is no known theory of a vector field
that would be the analogous of  the Galileon theories of a scalar,
i.e. theories for a vector field which would have field equations of
just second order on flat spacetime (besides of course Maxwell
theory). Similarly, as far as we know, there is no complete
classification of vector theories that would have field equations of
order less than or equal to two on flat spacetime\footnote{Notably,
Horndeski studied the most general tensor-vector theory with second
order field equations on arbitrary spacetimes (both for the metric and
vector) and charge conservation, with the additional assumption that
field equations reduce to Maxwell equations on flat space, see the
original Ref. \cite{Horndeski:1976gi} for more details.}. The purpose of
this work is to investigate these issues. We will prove in particular
that there is no theory equivalent to scalar Galileons for vector
fields, in any spacetime dimensions. 

This paper is organized as follows. In the following section we set some
notations and review some of the attempts towards building a ``vector
Galileon''. In Section \ref{sec:no-go-theorem}, we prove a no-go theorem
for vector Galileons, which is the major result of this paper. In
\ref{sec:furth-prop-equat}, we explore further properties of the
equations of motion and the action. We conclude in Section
\ref{sec:conc-diss}.

\section{Searching for vector Galileons, first attempts}
\label{sec:search-vect-galil}

Considering first a scalar field $\pi$, we recall that flat spacetime Galileon actions can be obtained for an arbitrary spacetime dimension $D$ by taking  
linear combinations of actions of the form 
\ba \label{LGAL1}
{\cal L}_{\npi} &=& 
{\mathcal A}_{(2\ndd+2)}^{\mu_{\vphantom{()}1}
\ldots \mu_{\vphantom{()}\ndd+1}\nu_{\vphantom{()}1}
\ldots \nu_{\vphantom{()}\ndd+1} }\pi_{,\mu_{n+1}} \pi_{,\nu_{n+1}} 
 \pi_{,\mu_{\vphantom{()}1} \nu_{\vphantom{()}1}} \ldots \pi_{,\mu_{\vphantom{()}\ndd} \nu_{\vphantom{()}\ndd}}
\ea
where $\npi$ indicates the number of times of $\pi$ occurs, $ \npi \equiv \ndd+2  \; (\geq 2)$, $N\leq  D+1,$ and the $2m$-contravariant tensor ${\mathcal A}_{(2m)}$ is defined by 
\ba \label{DEFAten}
{\mathcal{A}}_{(2m)}^{\mu_{\vphantom{()}1} \mu_{\vphantom{()}2}
\ldots \mu_{\vphantom{()}m} \nu_{\vphantom{()}1} \nu_{\vphantom{()}2}
\ldots \nu_{\vphantom{()}m}} \equiv
\frac{1}{(D-m)!}\,
\varepsilon^{\mu_{\vphantom{()}1}
\mu_{\vphantom{()}2}  \ldots
\mu_{\vphantom{()}m} \sigma_{\vphantom{()}1}\sigma_{\vphantom{()}2}\ldots
\sigma_{\vphantom{()}D-m}}_{\vphantom{\mu_{\vphantom{()}1}}}
\,\varepsilon^{\nu_{\vphantom{()}1} \nu_{\vphantom{()}2} \ldots
\nu_{\vphantom{()}m}}_{\hphantom{\nu_{\vphantom{()}1}
\nu_{\vphantom{()}2} \ldots
\nu_{\vphantom{()}2m}}\sigma_{\vphantom{()}1}
\sigma_{\vphantom{()}2}\ldots \sigma_{\vphantom{()}D-m}} \,.
\ea
where the (flat spacetime) totally antisymmetric Levi-Civita tensor is given by
\ba \label{DEFLC}
\varepsilon^{\mu_{\vphantom{()}1} \mu_{\vphantom{()}2} \ldots
\mu_{\vphantom{()}D}} \equiv - 
\delta^{[\mu_{\vphantom{()}1}}_1 \delta^{\mu_{\vphantom{()}2}}_2
\ldots \delta^{\mu_{\vphantom{()}D}]}_D \, 
\ea
with here and henceforth square brackets denoting unnormalized
permutations. The field equations obtained from the above Lagrangian
(\ref{LGAL1}) are purely second order equations (see
e.g. \cite{Deffayet:2009wt,Deffayet:2009mn}). This agrees in particular
with the counting of derivatives occurring in each of the Lagrangians
(\ref{LGAL1}). Indeed, for a fixed number of occurrences $N$ of the
scalar field, there are $2n+2 = 2(N-1)$ derivatives in the action, as
there should in order for the field equations to contain only second
derivatives. 

Let us now consider a set of $p$-forms $A^a_p$, where $a$ denotes the type of species, having field strength $F^a_{p+1} = d A^a_p$, such that the exterior derivative $d F^a$ vanishes. Motivated by the action above for a single field $\pi$, one can consider actions given by the formal expression  \cite{Deffayet:2010zh} 
\begin{equation}
{\cal L} = \varepsilon^{\mu_1 \mu_2 \dots}
\varepsilon^{\nu_1 \nu_2 \dots\dots}\,
F^a_{\mu_1\mu_2\dots} F^b_{\nu_1 \nu_2 \dots}
\left(\partial_{\mu_k} F^c_{\nu_l \nu_{l+1} \dots}\,\dots
\right)
\left(\partial_{\nu_j} F^d_{\mu_m \mu_{m+1}\dots}\,\dots\right) \, ,
\label{eq7}
\end{equation}
where the different species are labeled by $(a, b, \dots)$.
The number of indices contracted with the first and second
Levi-Civita tensors $\varepsilon$ must be the same and not greater than
$D$, but the two terms in parentheses may now involve {\it different}
species and therefore a {\it different} number of terms.  The Bianchi
identities (i.e.~$[d,d]=0$) ensure that only $\partial F$ 
appears in the field equations, which therefore remain of order two in derivatives. However, it was shown in \cite{Deffayet:2010zh}  that such actions are void when one considers just a single species of an odd-p form (i.e. in the expression above $p$ odd and all species indices $(a, b, \dots)$ taken to be the same). In this case indeed, the action is a total derivative and hence leads to vanishing field equations.
This is the case e.g. for the vector action in 5 dimensions given by \cite{Deffayet:2010zh}
\begin{eqnarray}
I&=&\int d^5 x\, \varepsilon^{\mu\nu\rho\sigma\tau}
\varepsilon^{\alpha\beta\gamma\delta\epsilon}\,
F_{\mu\nu} F_{\alpha\beta}\, \partial_\rho F_{\gamma\delta}\,
\partial_\epsilon F_{\sigma\tau}
\label{eq17}
\end{eqnarray}

A ``vector Galileon'', i.e. a vector theory with nonlinear second order field equation, if exists at all, has to be looked for in another way. One way to proceed is to consider the simplest possible case and assume that we start from an action that can be expressed in terms of the field strength $F_{\mu \nu} = \partial_{[\mu}A_{\nu]}$ of a vector $A_\mu$. If we demand that field equations are purely of second order and,  say, are polynomial functions of $\partial F$ of degree $N-1$, one sees that each such polynomial expression contains $3(N-1)$ space time indices (among which $2(N-1)$ are counting spacetime indices of derivatives, and $(N-1)$ indices are carried by the vector $A_\mu$), which must all but one be contracted together in order to yield field equations which have just one free vector indices (as it must be the case for the field equations of a vector $A_\mu$). Hence $3(N-1)$ must be odd which further means that the simplest possible such model (beyond the trivial Maxwell case, which corresponds to $N=2$) must have $N=4$, i.e. must have a Lagrangian of the form of linear combinations of monomial 
$A \partial F \partial F \partial F$, or rather, integrating by part, of the form 
\ba F F \partial F \partial F \label{genscavect}
\ea with all the 10 indices appearing there contracted together. In fact, the above construction (\ref{eq7}) of \cite{Deffayet:2010zh} yields such an action for $N=4$ (identical) vector field in $D=5$ (and this also can be generalized to higher dimensions) which reads as in (\ref{eq17}) but has vanishing field equations as we just said. It is however possible to scan all possible such models in $D=4$ dimensions by considering all the possible scalars of the form (\ref{genscavect})
 and see if there exists some combination of them yielding purely second order non trivial field equations. The corresponding detailed discussion is given in Appendix {\ref{appA}}, its outcome is that such a theory does not exist in $4$ dimensions. In the following, we will provide a more general proof of this, valid in particular for an arbitrary number of dimensions and an arbitrary power of the vector field.

\section{No-go theorem}
\label{sec:no-go-theorem}

In this section we prove the no-go theorem for generalized vector
Galileons on flat spacetime with arbitrary dimensions.

\subsection{Assumptions}
\label{sec:sum-goal-assump}

We consider a vector field $A_{\mu}$ in $D$-dimensional Minkowski
spacetime (${\cal M}$, $\eta_{\rho\sigma}$) and introduce the $U(1)$
gauge transformation 
\begin{equation}
 A_{\nu} \to \bar{A}_{\nu} = A_{\nu} + \lambda_{,\nu},
  \label{eqn:Abar}
\end{equation}
where $\lambda$ is an arbitrary scalar. We denote the equations of
motion for the vector $A_{\mu}$ by $E^{\mu}=0$.

Hereafter, we assume that (i) $E^{\mu}$ is invariant under the gauge
transformation (\ref{eqn:Abar}); (ii) $E^{\mu}$ depends on at most
second-order derivatives of $A_{\mu}$ as 
\begin{equation}
 E^{\mu}
  = E^{\mu}(A_{\nu}\,, A_{\nu,\rho}\,, A_{\nu,\rho\sigma}\,,
  \eta_{\rho\sigma}  \,, \epsilon) ~, \label{eqn:Emu}
\end{equation} 
and (iii) $E^{\mu}$ follows from variation of an action of the form  
\begin{equation}
 I = \int d^Dx \,{\cal L}(A_{\nu}\,, A_{\nu,\rho}\,, \cdots\,,
  A_{\nu,\rho_1\cdots\rho_h}\,, \cdots\,,\eta_{\rho\sigma}
  \,, \epsilon)
  \label{eqn:action}
\end{equation}
as
\begin{equation}
 \delta I = \int d^Dx E^{\mu}\delta A_{\mu}. \label{eqn:eom}
\end{equation} 
Here, $\epsilon$ represents the totally antisymmetric Levi-Civita tensor
given in \eqref{DEFLC}.

Under these assumptions we shall prove that $E^{\mu}$ is at most linear
in second derivatives of the vector field.

\subsection{From equations of motion to action}
\label{sec:from-equat-moti}

In this subsection we consider the condition under which equations of
motion of the form 
\begin{equation}
 E^{\mu}(A_{\nu}\,, A_{\nu,\rho}\,, A_{\nu,\rho\sigma}\,,
  \eta_{\rho\sigma}\,, \epsilon) 
\end{equation} 
can be derived from an action principle as in (\ref{eqn:eom}). For 
simplicity, we denote the above expression as $E^{\mu}(x)$ throughout this
subsection.

Since $E^{\mu}(x)$ can be written as
\begin{equation}
 E^{\mu}(x) = \int d^Dy E^{\mu}(y)\delta^D(y-x),
\end{equation}
the functional derivative of $E^{\mu}(x)$ w.r.t. $A_{\nu}(y)$ is
calculated as 
\begin{eqnarray}
 \frac{\delta E^{\mu}(x)}{\delta A_{\nu}(y)}
  & = & \frac{\partial E^{\mu}(y)}{\partial A_{\nu}(y)}\delta^D(y-x)
  - \frac{\partial}{\partial y^{\rho}}
  \left[\frac{\partial E^{\mu}(y)}{\partial A_{\nu,\rho}(y)}
   \delta^D(y-x)\right]  \nonumber\\
 & & 
  + \frac{\partial^2}{\partial y^{\rho}\partial y^{\sigma}}
  \left[\frac{\partial E^{\mu}(y)}{\partial A_{\nu,\rho\sigma}(y)}
   \delta^D(y-x)\right].
\end{eqnarray}
Multiplying this with a well-behaved function ${\cal F}(y)$ and
integrating over $y$, we obtain
\begin{equation}
\int d^Dy {\cal F}(y) \frac{\delta E^{\mu}(x)}{\delta A_{\nu}(y)} 
  = {\cal F}(x)\frac{\partial E^{\mu}(x)}{\partial A_{\nu}(x)} 
  + \frac{\partial {\cal F}(x)}{\partial x^{\rho}}
  \frac{\partial E^{\mu}(x)}{\partial A_{\nu,\rho}(x)}
  + \frac{\partial^2 {\cal F}(x)}{\partial x^{\rho}\partial x^{\sigma}}
  \frac{\partial E^{\mu}(x)}{\partial A_{\nu,\rho\sigma}(x)}.
\end{equation}
Similarly, we obtain
\begin{eqnarray}
\int d^Dy {\cal F}(y)\frac{\delta E^{\nu}(y)}{\delta A_{\mu}(x)}
 & = &
 {\cal F}(x)\frac{\partial E^{\nu}(x)}{\partial A_{\mu}(x)} 
  - \frac{\partial}{\partial x^{\rho}}
  \left[
   {\cal F}(x)\frac{\partial E^{\nu}(x)}{\partial A_{\mu,\rho}(x)}
  \right] \nonumber\\
 & & 
 + \frac{\partial^2}{\partial x^{\rho}\partial x^{\sigma}}
 \left[
   {\cal F}(x)\frac{\partial E^{\nu}(x)}{\partial A_{\mu,\rho\sigma}(x)}
 \right].
\end{eqnarray}
Therefore we obtain
\begin{eqnarray}
\int d^Dy {\cal F}(y)
 \left[
  \frac{\delta E^{\mu}(x)}{\delta A_{\nu}(y)} 
  - \frac{\delta E^{\nu}(y)}{\delta A_{\mu}(x)}
	       \right]
 & = & {\cal F}(x)
 [E^{\mu;\nu}(x)-E^{\nu;\mu}(x)] \nonumber\\
 & & +{\cal F}(x)
  \partial_{\rho}
  [ E^{\nu;\mu,\rho}(x) - \partial_{\sigma}E^{\nu;\mu,\rho\sigma}(x)] 
  \nonumber\\
 & & 
 + \frac{\partial {\cal F}(x)}{\partial x^{\rho}}
  [ E^{\nu;\mu,\rho}(x)+E^{\mu;\nu,\rho}(x) 
   - 2\partial_{\sigma}E^{\nu;\mu,\rho\sigma}(x)] \nonumber\\
 & & 
  + \frac{\partial^2 {\cal F}(x)}{\partial x^{\rho}\partial x^{\sigma}}
  [ E^{\mu;\nu,\rho\sigma}(x)-E^{\nu;\mu,\rho\sigma}(x)],
  \label{eqn:commutator}
\end{eqnarray}
where above and henceforth we adopt the notation
\begin{equation}
 T_{\cdots}^{\cdots;\nu,\rho\sigma} \equiv
  \frac{\partial T_{\cdots}^{\cdots}}{\partial A_{\nu,\rho\sigma}},
  \quad
 T_{\cdots}^{\cdots;\nu,\rho} \equiv
  \frac{\partial T_{\cdots}^{\cdots}}{\partial A_{\nu,\rho}},
  \quad
 T_{\cdots}^{\cdots;\nu} \equiv
  \frac{\partial T_{\cdots}^{\cdots}}{\partial A_{\nu}}.
  \label{eqn:notation}
\end{equation}

Let us now suppose that the equations of motion are derived from an action
principle as (\ref{eqn:eom}). In this case, the l.h.s. of
(\ref{eqn:commutator}) is written as 
\begin{equation}
\int d^Dy {\cal F}(y)
 \left[\frac{\delta}{\delta A_{\nu}(y)},
  \frac{\delta}{\delta A_{\mu}(x)}\right] I
\end{equation}
and thus should vanish for ${}^{\forall}{\cal F}$. By requiring that the
r.h.s. of (\ref{eqn:commutator}) vanish for ${}^{\forall}{\cal F}$, we thus
obtain the integrability conditions as
\begin{align}
  E^{\mu;\nu}-E^{\nu;\mu}
  + \partial_{\rho}
  (E^{\nu;\mu,\rho} - \partial_{\sigma}E^{\nu;\mu,\rho\sigma}) = 0~,
\end{align}
\begin{align}\label{eq:e2integ}
  E^{\nu;\mu,\rho}+E^{\mu;\nu,\rho}
   - 2\partial_{\sigma}E^{\nu;\mu,\rho\sigma}
  = 0~,
\end{align}
\begin{align} \label{eq:e1integ}
  E^{\mu;\nu,\rho\sigma}-E^{\nu;\mu,\rho\sigma} =  0~.
\end{align}

\subsection{Gauge invariance of equations of motion}
\label{sec:gauge-invar-equat}

We demand that $E^{\mu}$ is invariant under the $U(1)$ gauge
transformation:
\begin{equation}
E^{\mu}(\bar{A}_{\nu}\,, \bar{A}_{\nu,\rho}\,,
 \bar{A}_{\nu,\rho\sigma}\,,  \eta_{\rho\sigma}\,,  \epsilon)
= E^{\mu}(A_{\nu}\,, A_{\nu,\rho}\,, A_{\nu,\rho\sigma}\,,
  \eta_{\rho\sigma}\,,  \epsilon), \label{eqn:Emu-invariance}
\end{equation}
where $\bar{A}_{\mu}$ is given by (\ref{eqn:Abar}) and 
\begin{equation}
 \bar{A}_{\nu,\rho} = A_{\nu,\rho} + \lambda_{,\nu\rho}, \quad
 \bar{A}_{\nu,\rho\sigma} = A_{\nu,\rho\sigma}
 + \lambda_{,\nu\rho\sigma}. 
\end{equation}
By taking derivative of the gauge invariance equation
(\ref{eqn:Emu-invariance}) w.r.t. $\lambda_{,\nu\rho\sigma}$ and then
setting $\lambda=0$, we obtain the condition
\begin{equation}
 E^{\mu; \nu,\rho\sigma} + E^{\mu; \rho,\sigma\nu} + 
  E^{\mu; \sigma,\nu\rho} = 0. \label{eqn:invariance1}
\end{equation} 
In addition, it is obvious from the definition (\ref{eqn:notation}) that
\begin{equation}
  E^{\mu; \nu,\rho\sigma} = E^{\mu; \nu,\sigma\rho}.
   \label{eqn:commute1}
\end{equation}
Similarly by taking derivative of (\ref{eqn:Emu-invariance})
w.r.t. $\lambda_{,\nu\rho}$ and then setting $\lambda=0$, we obtain 
\begin{equation}
 E^{\mu; \nu,\rho} + E^{\mu; \rho,\nu} = 0. \label{eqn:invariance2}
\end{equation}
Finally, by taking derivative of (\ref{eqn:Emu-invariance})
w.r.t. $\lambda_{,\nu}$ and then setting $\lambda=0$, we obtain
\begin{equation} 
 E^{\mu; \nu} = 0. 
\label{eqn:noAmu}
\end{equation}
This in particular implies that 
\begin{equation}
 E^{\mu} = E^{\mu}(A_{\nu,\rho}\,, A_{\nu,\rho\sigma}\,,
  \eta_{\rho\sigma}\,,  \epsilon). \label{eqn:independence-on-Amu}
\end{equation}

\subsection{Symmetries of $E^{\mu;\nu,\rho\sigma}$}
\label{sec:symmetry-emu-nu}

Here we summarize properties of $E^{\mu;\nu,\rho\sigma}$ and derive a
few additional properties. We have already obtained conditions
\eqref{eq:e1integ}, \eqref{eqn:commute1} and \eqref{eqn:invariance1}: 
\begin{align} \label{eq:e1s1}
  E^{[\mu;\nu],\rho\sigma} =  0~,
\end{align}
\begin{equation}
  E^{\mu; \nu,[\rho\sigma]} = 0~,
   \label{eq:e1s2}
\end{equation}
\begin{equation}
 E^{\mu; (\nu,\rho\sigma)} = 0 ~, \label{eq:e1b1}
\end{equation} 
where here and in the following, parentheses around space-time indices
mean a normalized symmetrization, and vertical bars (such as in the
equation below)  around indices mean that these indices are omitted in
the symmetrization (or antisymmetrization).

From \eqref{eq:e1s1} and \eqref{eq:e1b1}, we obtain 
\begin{equation}
 E^{(\mu|; \nu,|\rho\sigma)} = 0 ~. \label{eq:e1b2}
\end{equation} 
From \eqref{eq:e1s2}, \eqref{eq:e1b1} and \eqref{eq:e1b2}, we obtain
\begin{align}
  E^{(\mu; \nu,\rho)\sigma} = 0 ~, \label{eq:e1b3}
\end{align}
since \eqref{eq:e1s2} implies the identity
\begin{equation}
 2E^{(\mu;\nu,\rho)\sigma} 
  = E^{\mu;(\nu,\rho\sigma)} + E^{\nu;(\rho,\mu\sigma)} +
  E^{\rho;(\mu,\nu\sigma)} - E^{(\mu|;\sigma,|\nu\rho)}
\end{equation}  
and each combination in the r.h.s. vanishes. 

Hence, symmetrization over any three indices of 
$E^{\mu;\nu,\rho\sigma}$ vanishes. From the above equations, we also have 
\begin{align}
    E^{\mu;\nu,\rho\sigma} - E^{\rho;\sigma,\mu\nu} = 0~, \label{eq:e1s3}
\end{align}
since \eqref{eq:e1s2} implies the identity
\begin{equation}
 2E^{(\mu;\nu),\rho\sigma} - 2E^{(\rho;\sigma),\mu\nu}
  = 3E^{\mu;(\nu,\rho\sigma)} + 3E^{\nu;(\mu,\rho\sigma)}
  - 3E^{(\rho|;\sigma,|\mu\nu)} - 3E^{(\sigma|;\rho,|\mu\nu)}
\end{equation} 
and each combination in the r.h.s. vanishes.

\subsection{Final step of the proof}

Note that for $E^{\mu;\nu_1,\rho_1\sigma_1;\nu_2,\rho_2\sigma_2}$, the
relations in the previous subsection are inherited with respect to each
layers of derivatives $\{  \nu_1,\rho_1\sigma_1 \}$ and 
$\{  \nu_2,\rho_2\sigma_2 \}$. Thus we have 
\begin{align}\label{eq:e2sym}
  E^{(\mu;\nu_1|,\rho_1\sigma_1;|\nu_2),\rho_2\sigma_2} = E^{\rho_1; \sigma_1, (\mu, \nu_1;\nu_2),\rho_2\sigma_2} = E^{(\nu_2|; \sigma_1, |\mu, \nu_1);\rho_1,\rho_2\sigma_2} = 0~,
\end{align}
where at the first equal sign we have used \eqref{eq:e1s3}; at the
second equal sign we have used \eqref{eq:e1s1}; and at the third equal
sign we have used \eqref{eq:e1b2}.

On the other hand, from \eqref{eq:e1s1}, 
$E^{\mu;\nu_1,\rho_1\sigma_1;\nu_2,\rho_2\sigma_2}$ 
is symmetric w.r.t. the $\{\mu, \nu_1\}$ indices and the 
$\{\mu, \nu_2\}$ indices and thus is already perfectly symmetric
w.r.t. the $\{\mu, \nu_1, \nu_2\}$ indices. Thus \eqref{eq:e2sym} is
simply 
\begin{align}
  E^{\mu;\nu_1,\rho_1\sigma_1;\nu_2,\rho_2\sigma_2} = 0~.
\end{align}
In other words, $E^\mu$ is at most linear in $A_{\alpha,\beta\gamma}$
and thus can be written as 
\begin{align} \label{eq:e01}
  E^{\mu} = E^{\mu; \alpha,\beta\gamma} A_{\alpha,\beta\gamma}  + K^\mu~, \mbox{~~where~~}
  E^{\mu; \alpha,\beta\gamma} = E^{\mu; \alpha,\beta\gamma}(A_{\nu,\rho}\,, \eta_{\rho\sigma}\,,  \epsilon)~,~
  K^\mu = K^\mu(A_{\nu,\rho}\,, \eta_{\rho\sigma}\,,  \epsilon)~.
\end{align}
Note that in even spacetime dimensions, one cannot construct $K^\mu$
using $A_{\nu,\rho}\,, \eta_{\rho\sigma}$ and $\epsilon$, because all
those tensors have even number of indices. Thus $K^\mu$ is nonzero only
in odd spacetime dimensions. 

Eq.~(\ref{eq:e01}) is the main result of the present paper, meaning that
the equations of motion are at most linear in the second derivatives of
the vector field. This concludes the proof of the no-go theorem for generalized vector
Galileons.

\section{Further properties of equations of motion and action}
\label{sec:furth-prop-equat}

In the previous section we already proved the no-go theorem for
generalized vector Galileons in $D$-dimensional flat spacetime. In this
section we shall discuss further properties of the equations of motion
and the corresponding action.

\subsection{Symmetries of $E^{\mu;\nu,\rho\sigma;\alpha\beta}$ and $K^\mu$}
\label{sec:sym4-e32}

Expanding the spacetime derivative $\partial_\sigma$ in \eqref{eq:e2integ} using chain
rule, and then taking derivative with respect to
$A_{\alpha,\beta\gamma}$, we obtain
\begin{equation}\label{eq:e32}
E^{\nu;\mu,\rho\gamma;\alpha,\beta}+E^{\nu;\mu,\rho\beta;\alpha,\gamma}-E^{\nu;\alpha,\beta\gamma;\mu,\rho}-E^{\mu;\alpha,\beta\gamma;\nu,\rho}=0\,.
\end{equation}
Inserting \eqref{eq:e01} into \eqref{eq:e2integ} and using
\eqref{eq:e32}, we get that 
\begin{align}
  K^{\mu;\nu,\rho} + K^{\nu;\mu,\rho} = 0~.
\end{align}
Also, from \eqref{eqn:invariance2}, we have
\begin{align}
  K^{\mu;\nu,\rho} + K^{\mu;\rho,\nu} = 0~.
\end{align}

The above two conditions also imply
\begin{align} \label{eq:k-asym}
  K^{\mu;\nu,\rho} + K^{\rho;\nu,\mu} = 0~.
\end{align}
Thus $K^{\mu;\nu,\rho}$ is totally anti-symmetric. This anti-symmetry is
inherited in $K^{\mu; \nu_1,\rho_1;\nu_2,\rho_2;\ldots ;\nu_n,\rho_n}$
in $D=2n+1$ dimensions, such that 
$K^{\mu;\nu_1,\rho_1;\nu_2,\rho_2;\ldots ;\nu_n,\rho_n}$ is also totally 
anti-symmetric.

\subsection{$E^{\mu}$ in terms of field strength}

It is convenient to introduce the field strength as usual
\begin{equation}
 F_{\alpha\beta} \equiv
  \partial_{\alpha}A_{\beta}-\partial_{\beta}A_{\alpha}.  
\end{equation}
By introducing symmetric tensors
\begin{eqnarray}
 S_{\alpha\beta} & \equiv &
  \partial_{\alpha}A_{\beta}+\partial_{\beta}A_{\alpha}, \nonumber\\
 S_{\alpha\beta\gamma} & \equiv &
  \partial_{\alpha}\partial_{\beta}A_{\gamma}
  +\partial_{\beta}\partial_{\gamma}A_{\alpha}
  +\partial_{\gamma}\partial_{\alpha}A_{\beta},
\end{eqnarray} 
we can express the first- and second-order derivatives of $A_{\alpha}$ in terms of $F_{\alpha\beta}$, $F_{\alpha\beta,\gamma}$,
$S_{\alpha\beta}$ and $S_{\alpha\beta\gamma}$ as
\begin{eqnarray}
 \partial_{\alpha}A_{\beta} & = &
  \frac{1}{2}(S_{\alpha\beta}+F_{\alpha\beta}), \nonumber\\
 \partial_{\alpha}\partial_{\beta}A_{\gamma} & = &
  \frac{1}{3}(S_{\alpha\beta\gamma}+F_{\beta\gamma,\alpha}
  -F_{\gamma\alpha,\beta}). 
\end{eqnarray} 
Let us then introduce
\begin{equation}
 \tilde{E}^{\mu}(S_{\nu\rho}\,, F_{\nu\rho}\,, 
  S_{\nu\rho\sigma}\,, F_{\nu\rho,\sigma}\,, \eta_{\rho\sigma}\,,  \epsilon) \equiv
 E^{\mu}(A_{\nu,\rho}\,, A_{\nu,\rho\sigma}\,,
  \eta_{\rho\sigma}\,,  \epsilon), 
\end{equation} 
where we have used (\ref{eqn:noAmu})-(\ref{eqn:independence-on-Amu}). It is easy to show that
(\ref{eqn:invariance2}) and (\ref{eqn:invariance1}) are rewritten as 
\begin{equation}
 \frac{\partial \tilde{E}^{\mu}}{\partial S_{\nu\rho}}=0, \quad
 \frac{\partial \tilde{E}^{\mu}}{\partial S_{\nu\rho\sigma}}=0.
\end{equation} 
Therefore, the equations of motion depend only on $F_{\alpha\beta}$,
$F_{\alpha\beta,\gamma}$, $\eta_{\alpha\beta}$ and $\epsilon$.

\subsection{Gauge invariance of the action}
\label{sec:gi-of-action}

We have assumed the gauge-invariance of the equations of motion and we
have also assumed that the equations of motion can be derived from an 
action principle. On the other hand, the gauge invariance of the
action follows from those two assumptions.

To see that, we note that Eq. \eqref{eq:e01} leads to 
\begin{align}
  \partial_\mu  E^\mu = E^{\mu;\nu,\rho\sigma ; \alpha,\beta} A_{\nu,\rho\sigma} A_{\alpha,\beta\mu} + K^{\mu; \alpha,\beta} A_{\alpha,\beta\mu} + E^{\mu; \alpha,\beta\gamma}A_{\alpha,\beta\gamma\mu}~.
\end{align}
All those three terms on the r.h.s. vanish, because of 
Eqs. \eqref{eq:e32}, \eqref{eq:k-asym} and \eqref{eq:e1b2}
respectively. In particular, upon using \eqref{eqn:invariance2},
\eqref{eq:e1s1}, \eqref{eq:e1s2} and \eqref{eq:e1s3}, the first term is
rewritten as
\begin{eqnarray}
E^{\mu;\nu,\rho\sigma ; \alpha,\beta} A_{\nu,\rho\sigma} A_{\alpha,\beta\mu} 
 & = & \frac{1}{4}
 \left(E^{(\mu|;\nu,\rho\sigma;\alpha,|\beta)}
  + E^{(\sigma|;\alpha,\beta\mu;\nu,|\rho)}
 \right) A_{\nu,\rho\sigma} A_{\alpha,\beta\mu}  \nonumber\\
 & = & \frac{1}{4}
 \left(-E^{(\mu|;\nu,\rho\sigma;|\beta),\alpha}
  + E^{\mu;\beta,\alpha(\sigma|;\nu,|\rho)}
 \right) A_{\nu,\rho\sigma} A_{\alpha,\beta\mu},
\end{eqnarray}
 and thus vanishes because of \eqref{eqn:invariance2}. With
 $\partial_\mu E^\mu=0$ and the variation of the action \eqref{eqn:eom},
 one finds that the action is gauge invariant.

\subsection{Chern-Simons term}
\label{sec:int-action}

Here we discuss integration from the equations of motion to obtain the
corresponding action. We separate the action into two parts,  
\begin{equation}
I = I_{M} + I_{CS}\,,
\end{equation}
where
\begin{equation}\label{eq:sgm}
\delta I_{M} = \int d^Dx \left(-\frac{2}{3}E^{\mu;\nu,\rho\sigma} F_{\nu\rho,\sigma}\right)\delta A_\mu\,,
\end{equation}
and
\begin{equation} \label{eq:scs}
\delta I_{CS} = \int d^Dx \,K^{\mu}\delta A_\mu\,.
\end{equation}

The first term $I_{M}$ corresponds to a generalization of the standard
Maxwell action, leading to equations of motion linear in second
derivatives of the vector field. For example, an action of the form 
\begin{align} \label{eq:conj-lag}
 \int d^Dx \,{\cal{L}}(F_{\mu\nu}, \eta_{\mu\nu}, \epsilon)~.
\end{align}
is of this type. We conjecture that (\ref{eq:conj-lag}) is the most
general action of this type, leaving its proof as a future work.

As for the second term $I_{CS}$, we have already noted that $K^{\mu}=0$
(and thus $I_{CS}=0$) in even dimensions. We now integrate
\eqref{eq:scs} explicitly in odd ($D=2n+1$) dimensions. The total
anti-symmetry of 
$K^{\mu; \nu_1,\rho_1;\nu_2,\rho_2;\ldots ;\nu_n,\rho_n}$ implies that 
\begin{equation} \label{eq:keps}
K^{\mu;\rho_1,\sigma_1;\cdots;\rho_n,\sigma_n} = K_0\epsilon^{\mu\rho_1\sigma_1 \rho_2 \sigma_2 \cdots \rho_n\sigma_n}\,,
\end{equation}
where $K_0$ is a constant\footnote{To show that $K_0$ is a constant, we note that 
$K^{\mu;\nu_1,\rho_1;\nu_2,\rho_2;\ldots ;\nu_n,\rho_n;\nu_{n+1},\rho_{n+1}}$ 
vanishes identically in $D=2n+1$ dimensions because of its total
anti-symmetry.}. We can then integrate this quantity, step by step, to
obtain: 
\begin{equation}
K^{\mu}= \kappa \,\epsilon^{\mu\rho_1\sigma_1 \rho_2 \sigma_2 \cdots \rho_n\sigma_n}F_{\rho_1\sigma_1}\cdots F_{\rho_n\sigma_n}\,,
\end{equation}
where $\kappa$ is a constant, and the integration constants at each step
are forced to be zero by the symmetries. Using this form, we can write
the variation of the action as
\begin{eqnarray}
\delta I_{CS} &=& \kappa \int d^Dx\,\epsilon^{\mu\rho_1\sigma_1 \rho_2 \sigma_2 \cdots \rho_n\sigma_n}F_{\rho_1\sigma_1}\cdots F_{\rho_n\sigma_n}\delta A_\mu\,,\nonumber\\
&=& 2^{(D-1)/2}\kappa \int d^Dx\,\epsilon^{\mu\rho_1\sigma_1 \rho_2 \sigma_2 \cdots \rho_n\sigma_n}\partial_{\rho_1}A_{\sigma_1}\cdots \partial_{\rho_n}A_{\sigma_n}\delta A_\mu\,,\nonumber\\
&=& 2^{(D-1)/2}\kappa  \int d^Dx\,\epsilon^{\mu\rho_1\sigma_1 \rho_2 \sigma_2 \cdots \rho_n\sigma_n}\partial_{\rho_1}\delta A_{\sigma_1}\cdots \partial_{\rho_n}A_{\sigma_n} A_\mu\,,
\end{eqnarray}
where in the last step, we integrated by parts and used the symmetry of
the Levi-Civita tensor. The relation between the last two lines allows
us to write: 
\begin{eqnarray}
\delta I_{CS} &=& \frac{2^{(D+1)/2}}{D+1} \kappa \int d^Dx\,\epsilon^{\mu\rho_1\sigma_1 \rho_2 \sigma_2 \cdots \rho_n\sigma_n}\left[ \partial_{\rho_1}\delta A_{\sigma_1}\cdots \partial_{\rho_n}A_{\sigma_n} A_\mu +\cdots+
\partial_{\rho_1}A_{\sigma_1}\cdots \partial_{\rho_n}\delta A_{\sigma_n} A_\mu \right.\nonumber\\
&&\left.\qquad\qquad\qquad\qquad\qquad\qquad\qquad\qquad+
\partial_{\rho_1}A_{\sigma_1}\cdots \partial_{\rho_n}A_{\sigma_n}\delta A_\mu \right]\nonumber\\
&=& \frac{2^{(D+1)/2}}{D+1} \kappa \int d^Dx\,\epsilon^{\mu\rho_1\sigma_1 \rho_2 \sigma_2 \cdots \rho_n\sigma_n} \delta \left[ \partial_{\rho_1}A_{\sigma_1}\cdots \partial_{\rho_n}A_{\sigma_n} A_\mu\right]\,,
\end{eqnarray}
which can be trivially integrated to give
\begin{equation}
{I_{CS} = \frac{2\,\kappa}{D+1} \int d^Dx\,\epsilon^{\mu\rho_1\sigma_1 \rho_2 \sigma_2 \cdots \rho_n\sigma_n} F_{\rho_1\sigma_1}\cdots F_{\rho_n \sigma_n} A_\mu}\,.
\end{equation}
This is precisely the Chern-Simons term in $D=2\,n+1$ dimensions.

\section{Conclusion and discussions}
\label{sec:conc-diss}

Given the assumptions of flat spacetime, a single copy of $U(1)$ gauge
field, gauge invariance and Lorentz invariance of equations of motion,
that the equations of motion are second order and follow from an action
principle, we have shown that the Galileon-like terms are not
allowed. The equations of motion are at most linear in 
$\partial\partial A$. 

Since we have assumed that the equations of motion follow from an action principle, it would be necessary, for completeness, to derive explicitly the form of the action. In Sec.\ref{sec:int-action} we showed that the terms containing no $\partial\partial A$ can be integrated into a Chern-Simons term. For what concerns the other piece of the field equations (the one linear in linear in $\partial\partial A$) it is clear that actions of the form (\ref{eq:conj-lag}) lead to such equations. 
However, the proof that this is the only possibility is left for a future work.

As typically in physics, a no-go theorem is no better than its assumptions. When we relax our assumptions, equations of motion nonlinear in $\partial\partial A$ can be obtained. For example, if multiple copies of $U(1)$, or a non-Abelian gauge field is considered, Galileon-like terms can appear \cite{Deffayet:2010zh} 
\begin{equation}
{\cal L} = \varepsilon^{\mu_1 \mu_2 \dots}
\varepsilon^{\nu_1 \nu_2 \dots\dots}\,
F^a_{\mu_1\mu_2\dots} F^b_{\nu_1 \nu_2 \dots}
\left(\partial_{\mu_k} F^c_{\nu_l \nu_{l+1} \dots}\,\dots
\right)
\left(\partial_{\nu_j} F^d_{\mu_m \mu_{m+1}\dots}\,\dots\right) \, .
\end{equation}
As another example, in non-flat spacetime, vector Horndeski term arise \cite{Horndeski:1976gi}
\begin{align}
{\cal{L}} \propto \sqrt{-g} \,\epsilon^{\alpha\beta\gamma\delta}\,\epsilon^{\mu\nu\rho\sigma}F_{\mu\nu} F_{\alpha\beta} R_{\rho\sigma\gamma\delta}~.
\end{align}
Note that similar actions can be obtained by ``covariantizing'' actions (i.e. in this context coupling to gravity, while maintaining the second order nature of the field equations, see the discussion in \cite{Deffayet:2010zh}) such as \eqref{eq17} which are trivial on flat space-times, but lead to a non trivial dynamics on curved manifolds. It would be interesting to probe systematically more general theories by relaxing our conditions.

\acknowledgments
This work was initiated during CD's visit to Kavli Institute for the
Physics and Mathematics of the Universe, University of Tokyo and
completed during CD and SM's visit to Yukawa Institute for Theoretical
Physics, Kyoto University (YITP workshop YITP-T-13-08 on ``modified
gravity''). We thank them for providing warm hospitality and stimulating
atmosphere during our visits. 
The research of CD leading to these results
has received funding from the European Research Council under the
European Communitys Seventh 
Framework Programme (FP7/2007-2013 Grant Agreement no. 307934). 
AEG acknowledges financial support from the European Research Council
under the European Union's Seventh Framework Programme (FP7/2007-2013) /
ERC Grant Agreement n. 306425 ``Challenging General Relativity''. 
SM acknowledges the support by Grant-in-Aid for Scientific Research
24540256 and 21111006, and by PICS collaboration programme. 
YW is supported by a Starting Grant of the European Research Council
(ERC STG grant 279617), and the Stephen Hawking Advanced Fellowship. 
The work of AEG, SM and YW was supported by WPI Initiative, MEXT,
Japan.

\begin{appendix}

\section{The simplest vector Galileon in 4 dimension: a no go theorem}
\label{appA}
We are looking for possible ``Galileons'' with one forms. I.e. we seek for an action for a spin 1 whose equations of motion are purely second order. Restricting ourselves to polynomial forms, as shown in the main text, the first non trivial possible Lagrangian must be a linear combination of the following 18 terms 
\ba
{\cal L}_1 &=& F_{\mu \nu} F^{\mu \nu} (\partial_\lambda F^{\rho \sigma})( \partial^\lambda F_{\rho \sigma}) , \;\;
{\cal L}_2 = F_{\mu \nu} F^{\mu \nu} (\partial_\rho F^{\lambda}_{\hphantom{\lambda}\sigma})( \partial_\lambda F^{\rho \sigma}) , \;\;
{\cal L}_3 = F_{\mu \sigma} F^{\mu \nu} (\partial^\rho F^{\lambda}_{\hphantom{\lambda}\rho})( \partial^\sigma F_{\nu \lambda}) \nonumber\\
{\cal L}_4 &=& F_{\mu \sigma} F^{\mu \nu} (\partial_\nu F^{\rho \lambda})( \partial^\sigma F_{\rho \lambda}),\;\;
{\cal L}_5 = F_{\mu \sigma} F^{\mu \nu} (\partial^\rho F_\nu^{\hphantom{\nu} \lambda})( \partial^\sigma F_{\rho\lambda}),\;\;
{\cal L}_6 = F_{\lambda \rho} F_{\mu \nu} (\partial^\mu F^\rho_{\hphantom{\rho} \sigma} )(\partial^\nu F^{\lambda \sigma}) \nonumber\\
{\cal L}_7 &=& F_{\lambda \rho} F_{\mu \nu} (\partial^\lambda F^\rho_{\hphantom{\rho} \sigma})( \partial^\mu F^{\nu \sigma} ),\;\;
{\cal L}_8 = F_{\lambda \rho} F_{\mu \nu} (\partial^\sigma F_\sigma^{\hphantom{\sigma} \rho})( \partial^\mu F^{\nu \lambda}) ,\;\;
{\cal L}_9 = F_{\lambda \rho} F_{\mu \nu} (\partial^\sigma F_\sigma^{\hphantom{\sigma} \nu})( \partial^\mu F^{\lambda \rho}) \nonumber \\
{\cal L}_{10} &=& F_{\lambda \rho} F_{\mu \nu} (\partial^\sigma F^{\lambda \rho})( \partial^\mu F_\sigma^{\hphantom{\sigma} \nu}),\;\;
{\cal L}_{11} = F_{\lambda \rho} F_{\mu \nu} (\partial^\sigma F^{\nu \rho} )(\partial^\mu F_\sigma^{\hphantom{\sigma} \lambda}) ,\;\;
{\cal L}_{12} = F^{\mu \nu} F_{\mu \nu} (\partial_\sigma F_\rho^{\hphantom{\rho} \sigma})( \partial_\lambda F^{\lambda \rho}) \nonumber\\
{\cal L}_{13} &=& F_{\mu}^{\hphantom{\mu} \sigma} F^{\mu \nu} (\partial^\rho F_{\nu \rho})( \partial^\lambda F_{\sigma \lambda}) ,\;\;
{\cal L}_{14} = F_{\mu}^{\hphantom{\mu} \sigma} F^{\mu \nu} (\partial_\lambda F_{\sigma}^{\hphantom{\sigma} \rho})(\partial^\lambda F_{\nu \rho}) ,\;\;
{\cal L}_{15} = F_{\mu}^{\hphantom{\mu} \sigma} F^{\mu \nu}  (\partial^\lambda F_{\nu \rho})( \partial^\rho F_{\sigma \lambda}) \nonumber \\
{\cal L}_{16} &=& F_{\lambda \rho} F_{\mu \nu} (\partial^\mu F^{\lambda \sigma})( \partial^\rho F^\nu_{\hphantom{\nu} \sigma}) ,\;\;
{\cal L}_{17} = F_{\lambda \rho} F_{\mu \nu} (\partial_\sigma F^{\lambda \rho} )(\partial^\sigma F^{\mu \nu}) ,\;\;
{\cal L}_{18} = F_{\lambda \rho} F_{\mu \nu} (\partial_\sigma F^{\nu \rho})( \partial^\sigma F^{\mu \lambda}) \label{eighteen}
\ea
which exhaust all possible index contractions inside terms with the structure of $F F \partial F \partial F$. 
Note that some of these terms can be seen to be proportional to each other using the Bianchi identities
\ba
\partial_{[\mu}F_{\nu \rho]} = 0.
\ea
We are looking for an action density, in $D=4$ dimensions, of the form 
\ba \label{MOST}
\sum_{i=1}^{i=18} {\cal C}_{i} {\cal L}_i.
\ea
The action density (\ref{eq17}), although written in 5 dimensions, can be expressed as a linear combination of the above form with the following coefficients\footnote{This rewriting is similar to the one which can be done for scalar Galileons with actions of the form (\ref{LGAL1}), which is just a linear combination of contractions of $n$ factor of $\partial \partial \pi$ with two $\partial \pi$} 
\begin{equation}
\begin{array}{llllll}
{\cal C}_1 = -4, & {\cal C}_2 = 0, & {\cal C}_3 = 32, & {\cal C}_4 =8, & {\cal C}_5 =0, &{\cal C}_6 =0 \\
{\cal C}_7 = 0,& {\cal C}_8 = 0,& {\cal C}_9 = 16, &{\cal C}_{10} =0,&{\cal C}_{11} =0, &{\cal C}_{12} =-8 \\
{\cal C}_{13} = -16,& {\cal C}_{14} = 16,& {\cal C}_{15} = 0, &{\cal C}_{16} =16,&{\cal C}_{17} =-4, &{\cal C}_{18} =0 
\end{array}\label{Cepsi}
\end{equation}
Let us now consider the most general form (\ref{MOST}). If we demand that the equations of motion do not contain fourth and third derivatives, we obtain (after quite long calculations) the 8 conditions
\begin{equation}
\begin{array}{clcl}
{\cal C}_1 =& \frac{1}{2} {\cal C}_{12} -\frac{1}{2} {\cal C}_2  &
{\cal C}_3 =& -\frac{1}{2} {\cal C}_{11} +\frac{3}{2} {\cal C}_{14} +\frac{3}{2} {\cal C}_{15}+ \frac{1}{2} {\cal C}_{16}-{\cal C}_{18}  \\
{\cal C}_4 =& -\frac{1}{4} {\cal C}_{11} +2 {\cal C}_{12} +\frac{3}{4} {\cal C}_{14}+ \frac{1}{4} {\cal C}_{15}-\frac{1}{4}{\cal C}_{16}-\frac{3}{2} {\cal C}_{18} + {\cal C}_9  &
{\cal C}_5 =& - 4 {\cal C}_{12} - {\cal C}_{14} + {\cal C}_{15} + {\cal C}_{16} + 2 {\cal C}_{18} - 2 {\cal C}_9  \\
{\cal C}_6 =& \frac{1}{2} {\cal C}_{11} - \frac{1}{2}{\cal C}_{14} -\frac{1}{2} {\cal C}_{15} +\frac{1}{2} {\cal C}_{16} +  {\cal C}_{18}   &
{\cal C}_7 =& -2 {\cal C}_{10} - \frac{1}{2}{\cal C}_{11} + 4{\cal C}_{12} +\frac{1}{2} {\cal C}_{14} + \frac{1}{2} {\cal C}_{15}+   \frac{1}{2} {\cal C}_{16}- 4 {\cal C}_{17} - {\cal C}_{18}  \\
{\cal C}_8 =&  \frac{1}{2}{\cal C}_{11} + 4 {\cal C}_{12} + \frac{1}{2}{\cal C}_{14}  + \frac{1}{2} {\cal C}_{15}-   \frac{1}{2} {\cal C}_{16}+ {\cal C}_{18} + 2 {\cal C}_9  &
{\cal C}_{13} =& - {\cal C}_{14} - {\cal C}_{15}
\end{array} \label{RELCOEF}
\end{equation}
One can see that those relations are fulfilled by coefficients (\ref{Cepsi}). 
In the most general case, when the above relations (\ref{RELCOEF}) are fulfilled, one is left with a 10 parameters\footnote{In fact some terms below are just zero by virtue of the Bianchi identity, some must be some total derivatives, so the action has less than 10 free parameters} family of Lagrangians given by 
\ba
4 {\cal L} &=& {\cal C}_2(-2 {\cal L}_1 + 4 {\cal L}_2) 
 +{\cal C}_9( 4 {\cal L}_4 - 8 {\cal L}_5 + 8 {\cal L}_8 + 4 {\cal L}_9)
 + {\cal C}_{10} (4 {\cal L}_{10} - 8 {\cal L}_7 ) \nonumber \\
&&+ {\cal C}_{11} (4 {\cal L}_{11} - 2 {\cal L}_3 - {\cal L}_4 + 2 {\cal L}_6 - 2 {\cal L}_7 + 2 {\cal L}_8) 
+ 2 {\cal C}_{12} ({\cal L}_{1} + 2 {\cal L}_{12} +4 {\cal L}_4 -8 {\cal L}_5 +8  {\cal L}_7 + 8 {\cal L}_8) \nonumber \\
&& + {\cal C}_{14} (-4 {\cal L}_{13} + 4 {\cal L}_{14} +6 {\cal L}_3 +3 {\cal L}_4 -4  {\cal L}_5 -2 {\cal L}_6 + 2 {\cal L}_7 + 2 {\cal L}_8) \nonumber \\&& 
+ {\cal C}_{15} (-4 {\cal L}_{13} + 4 {\cal L}_{15} +6 {\cal L}_3 +{\cal L}_4 + 4  {\cal L}_5 -2 {\cal L}_6 + 2 {\cal L}_7 + 2 {\cal L}_8) \nonumber \\
&& + {\cal C}_{16} (4 {\cal L}_{16} + 2{\cal L}_3 -{\cal L}_4 +4   {\cal L}_5 +2 {\cal L}_6 + 2 {\cal L}_7 - 2 {\cal L}_8) 
+ {\cal C}_{17} (4 {\cal L}_{17} -16 {\cal L}_7) \nonumber \\
&& + {\cal C}_{18} (4 {\cal L}_{18} -4 {\cal L}_3 - 6 {\cal L}_4 +8   {\cal L}_5 +4 {\cal L}_6 -4 {\cal L}_7 + 4 {\cal L}_8)
\ea
However, this action leads to vanishing field equations on flat spacetime.

\end{appendix}

\end{document}